\documentclass[aps,pra,twocolumn,twoside,a4paper,amsfonts,showpacs]{revtex4-1}

\usepackage[utf8x]{inputenc}
\usepackage[english]{babel}
\usepackage{hyperref}
\usepackage{blindtext}
\usepackage{bm}
\usepackage{amsmath,amsthm}
\usepackage{amssymb}
\usepackage{mathrsfs,dsfont,nicefrac}

\usepackage{epsfig}
\usepackage{graphicx}

\renewcommand{\.}{\hspace{-0.5ex}}
\renewcommand{\vec}[1]{\mathbf{#1}}
\renewcommand{\S}[1]{\mathcal{S}}
\newcommand{\E}[1]{\mathcal{E}}
\newcommand{\SE}[1]{\mathcal{SE}}
\newcommand{\bra}[1]{\langle#1|}
\newcommand{\ket}[1]{|#1\rangle}

\newcommand{\tr}{\text{Tr}}
\renewcommand{\Re}{\text{Re}}

\newcommand{\kom}[2]{\left[#1,#2\right]}

\newcommand{\tot}{\text{tot}}
\newcommand{\PT}{\text{PT}}
\newcommand{\n}{\overline{n}}

\newcommand{\sys}{\text{sys}}
\newcommand{\env}{\text{env}}

\newcommand{\red}{\text{red}}

\newcommand{\inter}{\text{int}}

\newcommand{\therm}{\text{therm}}

\begin{document}
\title{System-environment correlations and Non-Markovian dynamics}
\author{A.~Pernice}
\email{Ansgar.Pernice@tu-dresden.de}
\author{J.~Helm}
\author{W.T.~Strunz}
\affiliation{Institut f\"ur Theoretische Physik, Technische Universit\"at Dresden, D-01062 Dresden, Germany}

%
\begin{abstract}
We determine the total state dynamics of a dephasing open quantum system using
the standard environment of harmonic oscillators. 
Of particular interest are random unitary approaches to
the same reduced dynamics and 
system-environment correlations in the full model. 
Concentrating on a model with an at times
negative dephasing rate, the issue of ``non-Markovianity'' will also be
addressed. Crucially, given the quantum environment, the appearance of 
non-Markovian dynamics turns out
to be accompanied by a loss of system-environment correlations.
Depending on the initial purity of the qubit state, these system-environment
correlations may be purely classical over the whole relevant time scale, 
or there
may be intervals of genuine system-environment entanglement. In the latter
case, we see no obvious relation between the build-up or decay of these quantum 
correlations and ``Non-Markovianity''.

\end{abstract}





\maketitle

\section{Introduction}
In open quantum system dynamics one 
accounts for influences of the external environment
\cite{breuer+petruccione-2002, weiss-2008, gardiner+zoller-2002}.
Despite the unitary time-evolution of the total state of system plus
environment, the dynamics of the system itself will in general be
non-unitary. Growing correlations between the system of interest and
its surroundings lead to a decay of the initially present coherences.
This line of thought is at the heart of decoherence theory and is put
forward to explain the appearance of classical properties in quantum
systems \cite{joos+zeh-2003, zurek-2003, strunz-2002,
  schlosshauer-2008, braun-2001}. Decoherence in particular is of
relevance for quantum technologies trying to make use of the vast
computational potential forecast to applied quantum information
processing \cite{nielsen+chuang-2002}.

In the regime of weak system-environment coupling and short
environmental correlation times the dynamics of an open quantum system
may be described in terms of the Born-Markov approximation. The
corresponding Markov master equation then has a generator in Lindblad
form \cite{lindblad-1976, kossakowski-1972}. The time evolution of the
system depends on its present state only. Often, however, such an
approximation is not justified. Then, memory effects -- often
incorporated by means of integrals over the past \cite{nakajima-1958,
  zwanzig-1960} -- start to play an essential role. Yet, it is known
that for arbitrary bath correlation functions approximate and sometimes even exact time-local master
equations can be derived \cite{breuer+petruccione-2002, strunz+yu-2004}. 
An important
example is given by the exact master equation for a damped harmonic
oscillator bilinearly coupled to a bath of harmonic oscillators
\cite{haake+reibold-1985, hu+paz+zhang-1992, strunz+yu-2004}. 
Here, we focus on a
dephasing qubit as an open quantum system using the standard harmonic 
oscillator environment, whose exact time local master equation is known.

In more recent developments, the question of how to define and 
distinguish
``Markovian'' from ``non-Markovian'' dynamics from a local (system)
perspective was addressed. The analysis has been based both on a
single snapshot of the dynamics \cite{wolf+eisert+cubitt+cirac-2008}
and on the full time evolution of the open quantum system within a
certain time interval \cite{breuer+laine+piilo-2009,
  rivas+huelga+plenio-2010}. In the latter approach, memory effects
associated with non-Markovian dynamics are expected to cause temporary
increase in the distinguishability of states (in terms of trace
distance, e.g.  \cite{breuer+laine+piilo-2009}) and the dynamics may
no longer be divisible~\cite{rivas+huelga+plenio-2010}. Under Markovian
dynamics, on the other hand, the decay of distinguishability will be
monotonic throughout and divisibility will be ensured.

In this context 
the ``flow of quantum information'' from system to environment and 
back is an often employed, and certainly intuitive picture. 
However, without a proper conceptual and theoretical framework,
such a picture should be used with caution.

Care has also
to be taken with respect to the need for a proper (quantum)
environment. It should be noted that for single-qubit
decoherence the dynamics may always be described in terms of
\emph{stochastic} fluctuations of \emph{external fields}, i.e., the 
dynamics has a
random unitary representation
~\cite{kummerer+maassen-1987,landau+streater-1993,buscemi+chiribella+dariano-2005}.
The dynamics may thus be modeled
\emph{without} invoking a quantum environment at all. Higher dimensions then two
are needed to see decoherence that can only be understood in terms of
a proper quantum environment \cite{helm+strunz-2009}.

The role and nature of \emph{system-environment correlations} in open quantum system
dynamics has raised some notable interest lately. In the context of
quantum discord \cite{ollivier+zurek-2001}, for example, total states
with no quantum correlation (zero discord) were shown to be the most
general class of initial states allowing for completely positive (CP)
reduced dynamics \cite{rodriguez-rosario+al-2008,
  shabani+lidar-2009}. Another interesting line are investigations
about the relation
between decoherence and system-environment
entanglement. Here, it is possible that the system essentially decoheres
completely \emph{without} becoming entangled with its environment at all
\cite{eisert+plenio-2002, pernice+strunz-2011}. Such cases show that
classical system-environment correlations alone may account for a vast 
number of phenomena related to ``open quantum systems''. As we will
also see in this paper, often
an exchange of \emph{quantum} information between 
system and environment cannot be proven.

In order to shed light on the nature of system-environment correlations
in open system dynamics in a non-trivial (non-Lindblad) regime, 
and the possible relation to recent definitions of 
``Non-Markovianity''
we here
investigate single-qubit dephasing due to the coupling to an oscillator environment \cite{piilo+breuer-2011,pernice+strunz-2011, helm+strunz-2010}. In so doing,
we favour to investigate the dynamics and ``quantumness'' of
system-environment correlations thus avoiding the study of 
the somewhat vague notion of (quantum) ``information flow'' in open quantum 
system dynamics.

The paper is structured as follows: in Section \ref{sec2} we will present our model and give an exact, useful expression for the total system-environment state. Section \ref{sec:rand-unit-models} will be concerned with the derivation of a random unitary representation for the reduced qubit state at time $t$, showing that from the system perspective alone, no quantum environment is necessary to model the dynamics. The two inherently different approaches in Sections~\ref{sec2} and~\ref{sec:rand-unit-models} are exact and thus describe decoherence in the Markovian as well as the non-Markovian regime. Accounting for non-Markovianity, in Section~\ref{sec4} we concentrate on a super-ohmic spectral density which ensures an at times negative dephasing rate.
In Section~\ref{sec5} we employ a measure for system-environment correlations and relate it to ``non-Markovianity'' in the sense of~\cite{breuer+laine+piilo-2009,rivas+huelga+plenio-2010}. Following earlier work~\cite{pernice+strunz-2011}, in Section~\ref{sec6} we finally investigate the nature of these correlations. We find that for most qubit initial states there is no relation between
``non-Markovianity'' and the build-up or decay of \emph{quantum} correlations. 
Furthermore, even if there are time intervals where the total state of system 
and environment is entangled, there is no obvious relation to the
periods of ``non-Markovianity''.
We will draw our conclusions in Section~\ref{sec7}.
\section{Quantum decoherence model: reduced and full dynamics}\label{sec2}
Continuing our earlier work on the dynamics of system-environment correlations
for a dephasing qubit \cite{pernice+strunz-2011}, we start from a typical
model~\cite{feynman+vernon-1963,caldeira+leggett-1981} 
with total Hamiltonian
\begin{equation}\label{opensystemmodel}
H_\tot=H_\sys+H_\inter+H_\env,
\end{equation}
by coupling a qubit non-dissipatively to a bath of harmonic oscillators
through the choices
~\cite{skinner+hsu-1986,unruh-1995,kuang+zeng+tong-1999,yu+eberly-2002,yuan+kuang+liao-2010}
\begin{eqnarray}
H_\sys & = & \frac{\hbar \Omega}{2}\sigma_z \\ \nonumber 
H_{\rm int} & = & \sigma_z\otimes \sum_{\lambda=1}^N \hbar g_\lambda a_\lambda^\dagger+\text{h.c.}\\ \nonumber
H_\env & = & \sum_{\lambda=1}^N\hbar\omega_\lambda a_\lambda^\dagger a_\lambda.
\end{eqnarray}
Here $\Omega$ denotes the energy difference between the qubit states 
and the coefficients $g_\lambda$ describe the coupling
strengths between the qubit and each environmental mode of frequency
$\omega_\lambda$ and annihilation and creation operators $a_\lambda, 
a_\lambda^\dagger$. 

As environmental initial state we choose a thermal state
$\rho_\therm^\lambda=({\bar n}_\lambda + 1)^{-1}\exp[-\hbar\omega_\lambda a_\lambda^\dagger a_\lambda
/k_B T]$ for each oscillator with the mean thermal occupation number 
${\bar n}_\lambda = (\exp[\hbar\omega_\lambda/k_B T]-1)^{-1}$
at temperature $T$. Initially, we assume no system-environment 
correlations such that
the total initial state is simply given by the
product $\rho_\tot(0)=\rho_{\sys}\otimes\rho_\therm$.
Accordingly, for the reduced system state $\rho_\red(t)=\tr_\env[\rho_\tot(t)]$
the dynamical map ${\cal{E}} (t,0)$:
$\rho_{\red}(0) \rightarrow  \rho_{\red}(t)$ is completely positive (CP)
with $\rho_{\red}(0) = \rho_{\sys}$.

Already at this stage we emphasise that on the reduced level, this
dynamics can equally well be described by random unitary dynamics as
will be elaborated upon in Sec.~\ref{sec:rand-unit-models}.

The quantum dephasing model (\ref{opensystemmodel}) may be
solved without any approximation. A possible approach to the
time-local master equation for the system state is provided by the 
non-Markovian quantum state
diffusion approach to open systems \cite{diosi+gisin+strunz-1998,
  strunz+yu-2004}. We find for the reduced density operator
$\rho_\red(t)=\tr_\env[\rho_\tot(t)]$
\begin{equation}
\dot{\rho}_\red=-i\frac{\Omega}{2}\kom{\sigma_z}{\rho_\red}-\frac{\gamma(t)}{2}\left(\rho_\red-\sigma_z\,\rho_\red\,\sigma_z\right).
 \label{equ:masterequation-decoherence}
\end{equation}
This equation is solved by
\begin{equation}
  \rho_\red(t)=\begin{pmatrix}
    \rho_{00}&\mathcal{D}(t)\rho_{01}\\
    \mathcal{D}^*(t)\rho_{10}&\rho_{11}
\end{pmatrix},
\label{equ:solution-roh_red}
\end{equation}
with 
\begin{equation}\label{decoherence_factor}
\mathcal{D}(t)=\exp\left[-i\Omega t-\int_0^t\gamma(s)ds\right],
\end{equation}
and where the $\rho_{ij}$ represent the initial state of the qubit.

Equations~(\ref{equ:masterequation-decoherence}) and (\ref{equ:solution-roh_red}) involve
the time dependent dephasing rate $\gamma(t)$ which 
by means of the spectral density of the environment
$J(\omega)=\sum_{\lambda=0}^N|g_\lambda|^2\delta(\omega-\omega_\lambda)$
can be written as
\begin{equation}
\gamma(t)=4\int_0^t\.ds\int_0^\infty \.\.d\omega
J(\omega)\coth[\hbar\omega/2 k_B T]\cos[\omega s].
\label{equ:decoherence_rate}
\end{equation}
Later we will concentrate on environments that lead to periods in time with a {\it negative} dephasing rate.

Recently, we investigated system-environment correlations of this
model \cite{pernice+strunz-2011} and found the useful representation
\begin{equation}
  \rho_\tot(t)=\int\frac{d^2\xi}{\pi}\frac{1}{\n}e^{-|\xi|^2/\n}\;
  \hat{P}(t;\xi,\xi^*)\otimes\ket{\xi}\bra{\xi}
\label{equ:P-representation}
\end{equation}
of the total state. It represents a partial P-representation where the environmental degrees of freedom
are expanded in terms of coherent states $|\xi\rangle$. Here,
$\xi=(\xi_1,\xi_2,\cdots)$ is a vector of complex numbers and we consistently make use of
the notation $d^2\xi/\pi:=d^2\xi_1/\pi\; d^2\xi_2/\pi\cdots$ (see 
also~\cite{strunz-2005}). Furthermore, we symbolically write
$\exp[-|\xi|^2/\n]/\n:=\prod_\lambda\exp[-|\xi_\lambda|^2/\n_\lambda]/\n_\lambda$
involving the mean thermal occupation
number $\n_\lambda$ of the $\lambda$-th 
environmental mode.
The system part of the total state is encoded in a matrix-valued partial P-function
$\hat{P}(t)$ with values in the
$2\times2$ dimensional state space of the qubit. 

In order to represent a solution of the total Schr\"odinger-von-Neumann equation
with initial $\rho_\tot(0)=\rho_\sys\otimes\rho_\therm$, the partial P-function in
(\ref{equ:P-representation}) reads
\begin{equation}
\hat{P}(t;\xi,\xi^*)=
\begin{pmatrix}
\mathcal{A}^+(t;\xi,\xi^*)\rho_{00}&\mathcal{B}(t;\xi,\xi^*)\rho_{01}\\
\mathcal{B}^*(t;\xi,\xi^*)\rho_{10}&\mathcal{A}^-(t;\xi,\xi^*)\rho_{11}
\end{pmatrix}.
\label{equ:exact-solution}
\end{equation}
Here,
$\mathcal{A}^{\pm}=\exp[-A(t)\pm\left\{(a(t)|\xi)+(\xi|a(t))\right\}]$
and 
$\mathcal{B}=\exp[-i\Omega
t]\exp[B(t)-\left\{({b}(t)|\xi)-(\xi|b(t))\right\}]$,
where we have introduced the complex time dependent vectors
${a}(t)=(a_1(t),a_2(t),\cdots)$ and ${b}(t)$ with scalar product $(a(t)|\xi)\equiv\sum_\lambda a_\lambda^*(t)\xi_\lambda$
and vector components
\begin{eqnarray}
 a_\lambda(t)&=&\frac{1}{\n_\lambda}\int_0^t\left(g_\lambda e^{i\omega_\lambda s}\right)ds\\
 b_\lambda(t)&=&\frac{2\n_\lambda+1}{\n_\lambda}\int_0^t\left(g_\lambda e^{i\omega_\lambda s}\right)ds.
\end{eqnarray}
Furthermore, we use the abbreviations
\begin{eqnarray}
\nonumber
A(t)&=&2\,\Re\.\int_0^t\.ds\int_0^s\.d\tau\left[\sum_\lambda\frac{1}{\n_\lambda}|g_\lambda|^2e^{-i\omega_\lambda(t-s)}\right]\\
\nonumber
B(t)&=&2\,\Re\.\int_0^t\.ds\int_0^s\.d\tau\left[\sum_\lambda\frac{2\n_\lambda+1}{\n_\lambda}|g_\lambda|^2e^{-i\omega_\lambda(t-s)}\right].
\end{eqnarray}
Initially, $\hat P = \rho_\sys = \rho_\red(0)$ and
note that there are no approximations necessary to achieve the
result~(\ref{equ:exact-solution}) and thus via~(\ref{equ:P-representation}) to 
obtain the exact state of the composite system (see also \cite{pernice+strunz-2011}).

Later, we study the dynamics of system-environment correlations. Therefore,
a useful representation of the 
total state as in (\ref{equ:P-representation}) is of central importance. The 
local dynamics alone
is insufficient for the study of any quantities related to genuine open quantum
system dynamics, i.e., involving a proper quantum environment as in eq.~(\ref{opensystemmodel}). As we will elaborate upon
next, in our case the same reduced dynamics~(\ref{equ:masterequation-decoherence}) could have been obtained from a
stochastic Schrödinger dynamics,
not invoking a quantum environment at all.

\section{Random unitary representations}
\label{sec:rand-unit-models}

With an eye on experimental conditions,
decoherence of qubits is often modelled by random unitary dynamics~\cite{yu+eberly-2003,helm+strunz-2010}.
In terms of a dynamical map, this implies that there exists a relation
\begin{equation}\label{random_unitary}
\rho_{\red}(t) = \sum_k p_k U_k \rho_{\red}(0)U_k^\dagger
\end{equation}
with suitably chosen probabilities $p_k>0$ and unitary maps $U_k$.
Indeed, on the level of the reduced state, single-qubit decoherence
(and indeed, all single-qubit unital CP maps) can always be modelled in 
this way
~\cite{kummerer+maassen-1987,landau+streater-1993,buscemi+chiribella+dariano-2005}.
Thus, from a reduced point of view no quantum environment as in eq.~(\ref{opensystemmodel}) is required.
The reduced dynamics can be obtained from a local Schrödinger equation driven by a random Hermitian Hamiltonian.
By contrast, genuine \emph{quantum decoherence}
may be found in two-qubit systems~\cite{helm+strunz-2009}.
It is also worth noting that random unitary dynamics emerging from an
open quantum system with environmental initial pure state can always
be ``undone''(quantum error correction) \cite{gregoratti+werner-2003,
  trendelkamp-schroer+helm+strunz-2011}.

The most straightforward random unitary realization of single-qubit decoherence with
state (\ref{equ:solution-roh_red}) at time $t$ is provided by the simple quantum operation
\begin{eqnarray}\label{dephasing_process}
  \rho_{\red}(t) & = &
  \left(\frac{1+|\mathcal{D}(t)|}{2}\right) e^{-i\frac{\Omega}{2} t\sigma_z} \rho_{\red}(0) e^{i\frac{\Omega}{2} t\sigma_z}\\
  \nonumber
  & & +
  \left(\frac{1-|\mathcal{D}(t)|}{2}\right) e^{-i\frac{\Omega}{2} t\sigma_z}\sigma_z \rho_{\red}(0)\sigma_z e^{i\frac{\Omega}{2} t\sigma_z}
\end{eqnarray}
which is obviously of the form (\ref{random_unitary}) employing just
two unitaries $U_1 = \exp[-i\Omega t\sigma_z/2]$, $U_2 =
\exp[-i\Omega t\sigma_z/2] \sigma_z$ and probabilities $p_{1,2}=
(1\pm|\mathcal{D}(t)|)/2$.  Recall that according to
(\ref{decoherence_factor}), $|\mathcal{D}(t)|=\exp[-\int_0^t
  \gamma(s)ds]$.

It is worth noting that the very same formal relation holds true for the two-time map
\begin{equation}\label{equ:two_time_map}
{\cal E}(t,t'): \rho_{\red}(t') \rightarrow \rho_{\red}(t)
 \end{equation}
 such that
\begin{equation}\label{two-time_dephasing_process}
\begin{split}
  \rho_{\red}(t)&=\left(\frac{1+|\mathcal{D}(t,t')|}{2}\right) U_1(t-t')\rho_\red(t')U_1^\dagger(t-t')\\
  &\quad+\left(\frac{1-|\mathcal{D}(t,t')|}{2}\right)U_2(t-t')\rho_\red(t')U_2^\dagger(t-t')
  \end{split}
\end{equation}
with $|\mathcal{D}(t,t')|=\exp[-\int_{t'}^t \gamma(s)ds]$. However, as
$\gamma(s)$ need not be positive for all times (see later), the
prefactor of the second contribution,
$\frac{1-|\mathcal{D}(t,t')|}{2}$, may turn negative for $t'$ and $t$
near times of negative $\gamma(s)$. Thus, for such times $(t',t)$, the
map ${\cal E}(t,t')$ in the form (\ref{two-time_dephasing_process})
ceases to take the form of a random unitary map. Indeed, using the
Jamiolkowski isomorphism \cite{jamiolkowski-1972} it is
straightforward to see that $|\mathcal{D}(t,t')|<1$ or
\begin{equation}\label{equ:int_gamma}
\int_{t'}^t \gamma(s)ds > 0
\end{equation}
is a necessary and sufficient condition for the dephasing map ${\cal
  E}(t,t')$ defined above in eq.~(\ref{equ:two_time_map}) to be CP.

The random unitary form (\ref{dephasing_process}) of the dephasing map
${\cal E}(t,0)$ is simple but has the drawback of not representing an
intuitive dynamical picture of the process due to the time dependence
of the probabilities $p_1=p_1(t)$ and $p_2=p_2(t)$.  Here, we want to
develop an alternative random unitary representation in a more
systematic way that opens the door for further generalizations as
explained below.

Recall that pure dephasing of an open quantum system 
in the basis $\{|n\rangle\}$
implies a controlled-unitary form $U_{\rm tot}(t)=e^{-i H_{\rm
    tot}t/\hbar} = \sum_n |n\rangle\langle n|\otimes U_n(t)$ of the
total propagator \cite{helm+strunz+rietzler+wuerflinger-2011}. Here,
$U_n(t)=e^{-i H_n t /\hbar}$ with $H_n = \langle n|H_{\rm
  tot}|n\rangle$ are system-state-dependent propagators for the
environment (see also \cite{gorin+prosen+seligman+strunz-2004}). Pure
dephasing is then given by the dynamics $\langle n|\rho_{\rm
  red}(t)|m\rangle =\,$ $\tr_{\rm env}[U_n(t)\rho_{\rm
  env}(0)U_m^\dagger(t)]\cdot\langle n|\rho_{\rm red}(0)|m\rangle$.
In our case of a single qubit there is a single decoherence factor
$\mathcal{D}(t) = $\tr$_{\rm env}[U_0(t)\rho_{\rm env}(0)U_1^\dagger]$
as in (\ref{decoherence_factor}).  The two propagators are determined
by the environment Hamiltonians $H_i= \langle i |H_{\rm tot}|i\rangle$
with $H_1=\hbar\Omega/2 + \sum_{\lambda=1}^N \hbar g_\lambda
(a_\lambda^\dagger+a_\lambda) + \sum_{\lambda=1}^N\hbar\omega_\lambda
a_\lambda^\dagger a_\lambda$ and two sign changes for $H_0$.  We next
employ the Wigner representation of the environmental initial state
$W_0(\alpha,\alpha^*)=\int\frac{d^2\xi}{\pi}\, e^{\xi^*\alpha -
  \xi\alpha^*} \tr[e^{\xi a^\dagger -\xi^* a}\rho_{\rm
    env}]$ and for the operator $U_1^\dagger(t) U_0(t)$,
accordingly. The latter's Wigner Weyl symbol we denote by
$U_{01}(\alpha,\alpha^*)=\int\frac{d^2\xi}{\pi}\, e^{\xi^*\alpha -
  \xi\alpha^*} \tr\left[e^{\xi a^\dagger -\xi^* a}
  U_1^\dagger(t) U_0(t) \right]$.  We find
\begin{equation}\label{dephasing}
{\mathcal{D}(t)}=\int\frac{d^2\alpha}{\pi}\; W_0(\alpha,\alpha^*)\; U_{01}(\alpha,\alpha^*,t).
\end{equation}
Due to the harmonic properties of the environment, the corresponding
propagators $U_n(t)$ are known explicitly and lead to the phase factor
$U_{01}(\alpha,\alpha^*,t)=\exp[-i\Phi(\alpha,\alpha^*,t)]$ with the
phase
\begin{equation}
  \Phi(\alpha,\alpha^*,t)=\frac{\Omega}{2} t -2\, \sum_\lambda g_\lambda
  \alpha_\lambda\int_0^t e^{-i\omega_\lambda s}ds +\text{c.c}.
\end{equation}
We see that $\mathcal{D}(t)$ is just an average over a random complex
number of unit norm.  For a thermal initial state the initial Wigner
function $W_0=\frac{1}{{\bar n}+\frac{1}{2}} \exp[-|\alpha|^2/({\bar
  n}+\frac{1}{2})] $ is positive. Thus, expression (\ref{dephasing})
leads to a random unitary representation for ${\cal{E}}(t,0)$:
\begin{equation}\label{random_unitary_2}
\rho_{\rm red}(t)=\int\frac{d^2\alpha}{\pi}\; W_0(\alpha,\alpha^*)\; U_{\alpha}(t)\,
\rho_{\rm red}(0)\, U_{\alpha}^\dagger(t).
\end{equation}
This corresponds to a random unitary evolution of the qubit with
$U_{\alpha}=\exp[-i\int_0^t H_{\alpha}(s)ds/\hbar]$
and diagonal random Hamiltonian 
$H_{\alpha}(t)=\sigma_z\left(\frac{\hbar\Omega}{2} -\sum_\lambda g_\lambda (
\alpha_\lambda\int_0^t e^{-i\omega_\lambda s}ds + 
\alpha^*_\lambda\int_0^t e^{i\omega_\lambda s}ds\right).$ Note that in this representation
the probability of occurrence of a particular unitary evolution is given by the 
value of the initial
Wigner distribution and is thus time independent. 

The second random unitary representation  (\ref{random_unitary_2}) of the reduced dynamics reflects 
an ensemble of experiments where the unitary system dynamics is determined by the random $H_{\alpha}(t)$, driven by some (classical) 
stochastic process.  No quantum environment is involved.

Note that both random unitary representations of the dynamical
map ${\cal E}(t,0)$ are exact -- no restriction on the sign of $\gamma(s)$ is necessary.

It may appear tempting to {\it define} a two-time map ${\cal F}(t,t')$
through (\ref{random_unitary_2}) with $U_{\alpha}\rightarrow
U_{\alpha}(t,t')=\exp[-i\int_{t'}^t
H_{\alpha}(s)ds/\hbar]$. However, it is clear that ${\cal
  F}(t,t')\neq {\cal E}(t,t')$ unless $t'=0$. Indeed, while ${\cal F}(t,t')$ is a CP map for all $t,t'$ this ceases to be true for ${\cal E}(t,t')$ (see the next Section).

We close this section by pointing out an interesting additional
observation: the random unitary representation
(\ref{random_unitary_2}) for the quantum dephasing model is {\it not}
restricted to single-qubit-dephasing. In fact, for an arbitrary system
Hilbert space dimension, the very same construction works for all
dephasing factors ${\mathcal D}_{nm}(t)=\tr_{\rm env}[U_n(t)\rho_{\rm
  env}(0)U_m^\dagger]$ of a quantum oscillator environment model.  So
even for larger Hilbert space dimension than two -- on a local
level -- pure dephasing based on a quantum oscillator model like
(\ref{opensystemmodel}) {\it cannot} be distinguished from random
unitary dynamics. For genuine quantum decoherence, one needs ``more
quantum mechanical'' environments ~\cite{helm+strunz-2009}.

\section{Negative dephasing rate and "non-Markovianity"}\label{sec4}

The physics of the harmonic oscillator environment model is encoded in
its spectral density $J(\omega)$. As we are here interested in
instances of negative dephasing rate, we choose a particular super-ohmic
spectral density with sharp cutoff at frequency $\omega_c$
\begin{equation}\label{spectraldensity}
  J(\omega) = \kappa\frac{\omega^3}{\omega_c^2}\Theta(\omega-\omega_c)
\end{equation}
with $\Theta(\omega)$ the Heaviside step function and $\kappa$ a
dimensionless coupling constant.  In the high-temperature limit $k_B T
\gg \hbar\omega_c$ the
time dependent dephasing rate (\ref{equ:decoherence_rate}) can easily
be evaluated analytically, we get
\begin{equation}
\hbar\gamma(t)=8\kappa k_B T\left(\frac{\sin(\omega_c t)}{(\omega_ct)^2}
-\frac{\cos(\omega_c t)}{(\omega_ct)}\right).
\end{equation}

Though many of our results do not rely on any special choice of $J(\omega)$, in the following, whenever we show figures, we will use
the spectral density (\ref{spectraldensity}) and account for the high-temperature limit by choosing $T = 10\,\hbar\omega_c/k_B $. Furthermore, we choose $\kappa=10^{-2}$ throughout this paper.

Single qubit decoherence with negative dephasing rate is interesting in connection with two recently 
proposed ``measures of non-Markovianity''~\cite{breuer+laine+piilo-2009,rivas+huelga+plenio-2010}. 
In both definitions the notion of divisibility, related to the decomposition 
of the dynamical map according to 
$\mathcal{E}(t,t')=\mathcal{E}(t,t'')\mathcal{E}(t'',t')$ with $t\ge t''\ge t'$ is at the heart of ``non-Markovianity''. As explained 
around eq. (\ref{equ:int_gamma}), our $\mathcal{E}(t,t')$ ceases to be a 
CP map for time intervals of negative $\gamma(t)$. Indeed, as can be confirmed easily, for single qubit dephasing the two measures of non-Markovianity 
from ~\cite{breuer+laine+piilo-2009,rivas+huelga+plenio-2010}
are non-zero 
whenever $\gamma(t')<0$ for some $0<t'<t$. 

The fact that the map $\mathcal{E}(t,t')$ is no longer CP at times $t'>0$ is a consequence 
of growing correlations between system and environment. These correlations may be due to entanglement, but they need not 
be as will be shown in Section~\ref{sec6}.

\begin{figure}
  \includegraphics[width=\linewidth]{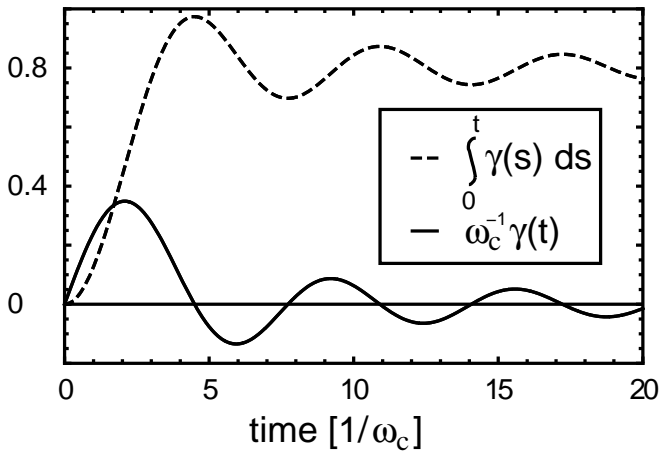}
  \caption{For our choice of super-ohmic spectral density, the dephasing rate $\gamma(t)$ shows
    domains of negative values. The integrated quantity
    $\int_0^t\gamma(s)ds$ stays positive, reflecting complete positivity
    of the dynamical map from $0$ to $t$.}
  \label{fig:figure1}
\end{figure}

As can be seen in Fig.~\ref{fig:figure1} our $\gamma(t)$ turns negative
in certain restricted periods, while the integral
\begin{equation}\label{equ:intgamma}
\int_0^t\gamma(s)ds = \frac{8k_B T}{\hbar\omega_c}
\left(1-\frac{\sin(\omega_c t)}{(\omega_ct)}\right)
\end{equation}
stays positive, as expected for the CP map $\mathcal{E}(t,0)$.

\section{Total state and system-environment correlations}\label{sec5}

Coupling to an environment leads to the build-up of correlations
between system and environment and thus to changes in local
entropies. From the point of view of information theory, the
reduced dynamics is regarded as a ``channel'' for quantum information. In this context, several
quantities related to von Neumann entropy $S=-\tr(\rho\log\rho)$ are of interest: e.g. $C_S(t)=S_{\sys}+S_{\env}-S_{\tot}$ as a measure for system-environment correlations \cite{holevo-2000}.
These quantities are hard to compute, unless one deals with very small systems or Gaussian
states.

Here we choose purity P=tr$(\rho^2)$ as an indicator
for the mixedness of states, which is related to the ``linear entropy'' via $S_L=1-P$. Clearly, as with entropy,
total purity $P_{\rm tot}$ is preserved under unitary evolution with $H_{\rm tot}$.
The sum of local purities $P_{\rm sys}(t)$, $P_{\rm env}(t)$ however, will be smaller
as $t>0$. For the initial product state we have $P_{\rm tot}=P_{\rm sys} P_{\rm env}$ 
and it appears natural for all $t\ge 0$ to consider the difference of logarithms of $P$
as a simple measure of correlations
\begin{equation}\label{correlation}
C = \log(P_{\rm tot}) - \log(P_{\rm sys}) - \log(P_{\rm env}).
\end{equation}
$C$ is easier to compute than $C_S$, but still one finds $C=C_S=0$ for uncorrelated states and $C=C_S=2\ln N$ for maximally 
entangled bipartite pure states of equal dimension $N$.
The dynamics of system-environment correlations is given by
\begin{eqnarray}\nonumber
 C(t) & = & \log\left(\frac{P_{\rm sys}(0)}{P_{\rm sys}(t)}\right)
     + \log\left(\frac{P_{\rm env}(0)}{P_{\rm env}(t)}\right)\\
     \label{correlation}
     &\equiv& C_\sys(t)+C_\env(t).
\end{eqnarray}
Here the contributions $C_\sys$ and $C_\env$ correspond to the amount of correlations created between system and environment as indicated by the increase of 
the local entropies in the two subsystems.

Having the total state (\ref{equ:P-representation}) at hand, all these quantities
can be determined easily for our dephasing qubit.
For instance, the qubit purity is readily determined to give
\begin{equation}\label{puritysystem}
P_{\rm sys}(t) = \frac{1}{2}\left(1+z^2 + (x^2+y^2)|{\mathcal{D}}(t)|^2\right).
\end{equation}
Here and in the following we denote by $\vec r =(x,y,z)=\tr[\vec\sigma\rho]$ the
coordinates of the Bloch vector of the {\it initial state of the qubit}.
Somewhat more involved, yet still easy to determine is the purity of the environment.
We find
\begin{equation}\label{purityenvironment}
P_{\rm env}(t) = \frac{1}{2}\left(1+z^2 + (1-z^2)|{\mathcal{G}}(t)|^2\right)P_{\rm env}(0)
\end{equation}
with the initial environmental purity
\begin{equation}
\label{initialpenv}
\log P_{\rm env}(0)=\int_0^\infty\.\.d\omega J(\omega)\log(\tanh(\hbar\omega/k_B T)).
\end{equation}
In (\ref{puritysystem}), the time dependence arises from the decoherence factor
$|{\mathcal{D}}(t)|=\exp[-\int_0^t\gamma(s)ds]$ of qubit dephasing
with the rate
$\gamma(t)=4\int_0^t\.ds\int_0^\infty \.\.d\omega
J(\omega)\coth[\hbar\omega/2 k_B T]\cos[\omega s]$ from
(\ref{equ:decoherence_rate}).
By contrast, for the environment the time dependence is governed by a factor
$|{\mathcal{G}}(t)|=\exp[-\int_0^t\Gamma(s)ds]$ with a dual rate
$\Gamma(t)=4\int_0^t\.ds\int_0^\infty \.\.d\omega
J(\omega)\tanh[\hbar\omega/2 k_B T]\cos[\omega s]$.

The rate of change of the correlation $C(t)$ from
(\ref{correlation}) stems from the two contributions $\dot C = \dot C_{\rm sys} + \dot C_{\rm env}$
with
\begin{equation}
\dot C_{\rm sys}(t) = \frac{2\gamma(t)}{a|{\mathcal D}(t)|^2+1}
\label{equ:Csys}
\end{equation}
and
\begin{equation}
\dot C_{\rm env}(t) = \frac{2\Gamma(t)}{b|{\mathcal G}(t)|^2+1},
\label{equ:Cenv}
\end{equation}
where the initial state of the qubit determines the factors $a=(1+z^2)/(1-z^2)$ and 
$b=(1+z^2)/(x^2+y^2)$.

Eqs. (\ref{equ:Csys}) and (\ref{equ:Cenv}) reflect a first important result: System and environment become more
correlated for $\gamma(t),\Gamma(t)>0$. More interestingly, system-environment correlations \emph{decrease} for negative
dephasing rates. In other words, during ``non-Markovian'' periods system and environment recover some of their initial independence.
As we will elaborate in the next section, these system-environment correlations may well be purely classical without the build-up
of entanglement.
\begin{figure}
  \includegraphics[width=\linewidth]{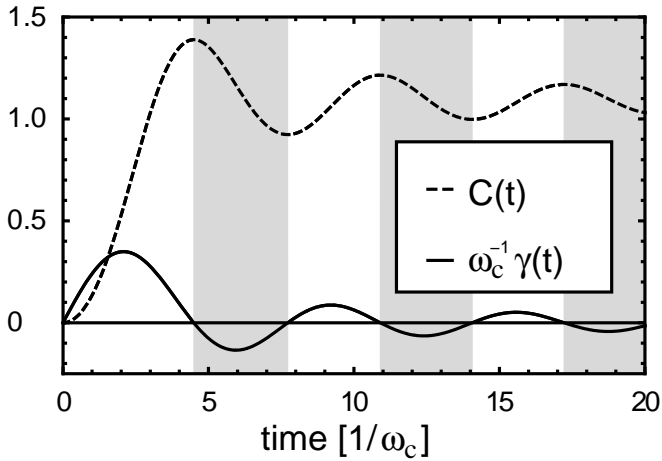}
  \caption{Dephasing rate $\gamma(t)$ (solid) and system-environment
    correlation $C(t)$ (dashed) against time for a qubit with initial purity $r=0.98$. While correlations grow for positive
dephasing rates ($\gamma(t)>0$), they decrease for
    $\gamma(t) < 0$ (highlighted domains).}
  \label{abb:figure2}
\end{figure}

In the high temperature limit considered here, by means of~(\ref{equ:Csys}) and~(\ref{equ:Cenv}) all quantities in~(\ref{correlation}) can 
be obtained readily. 
Reflecting the huge dimension of the environmental Hilbert space, it turns out that its contribution $C_\env(t)$ is
small compared to $C_\sys(t)$. Therefore, from~(\ref{equ:Csys}) we expect $\dot C(t)\sim\gamma(t)$.
In Fig.~\ref{abb:figure2} we display system-environment correlations $C(t)$ and dephasing rate $\gamma(t)$. Clearly,
changes in $C$ correlate with the
sign of $\gamma$, and thus $C(t)$ decreases in domains of non-Markovianity. As we will explain in the next Section, in this
case the total state is not entangled and thus $C(t)$ reflects classical correlations only.

Recall that this connection between system-environment correlations and non-Markovianity can only be established on the basis
of the total quantum state (\ref{equ:P-representation}).
For the random unitary representation (\ref{random_unitary_2}) of the same reduced dynamics, due to
the lack of an environment, the notion of system-environment correlations ceases to make sense.

\section{Quantum and classical system-environment correlations}\label{sec6}

Having access to the total state we can also investigate the nature of system-environment correlations.
In earlier work we have shown that quantum correlations need not exist in such
open system models, in particular in the high-temperature limit. In such a case,
the total state may still be separable and thus all correlations could be established using classical communication.
Here we argue very much as in \cite{pernice+strunz-2011}.

With the time and temperature dependent function
\begin{equation}
\begin{split}
 S(T,t)&=4 \int_0^t\.ds\int_0^s\.d\tau\int_0^\infty\.\.d\omega\\
 &\qquad\times\,J(\omega)\;\exp\left[\hbar\omega/kT\right]\cos[\omega(s-\tau)]
\end{split}
\label{equ:S_t_continuum}
\end{equation}
we have shown in~\cite{pernice+strunz-2011} that the total state is separable, as long as

\begin{equation}\label{equ:condition_sep}
S(T,t)\le\ln\sqrt{\frac{1-z^2}{x^2+y^2}}.
\end{equation}
In the high temperature limit, with our special choice of $J(\omega)$  this quantity can be easily evaluated yielding
\begin{equation}
S(T,t)=4\kappa\left(\frac12-\frac{\sin[\omega_c t]}{\omega_c t}-\frac{\cos[\omega_c t]}{(\omega_c t)^2}+\frac{1}{(\omega_c t)^2}\right).
\label{equ:S_special_choice}
\end{equation}
With criterion (\ref{equ:condition_sep}) we can indeed prove that the total state underlying
the correlation displayed in Fig.~\ref{abb:figure2} is separable. 

These findings show that the existence of a quantum environment does not imply 
(growing) entanglement. Moreover,
there is no connection between the non-Markovian character of the dynamics 
and the nature of system-environment correlations.

\begin{figure}
  \includegraphics[width=\linewidth]{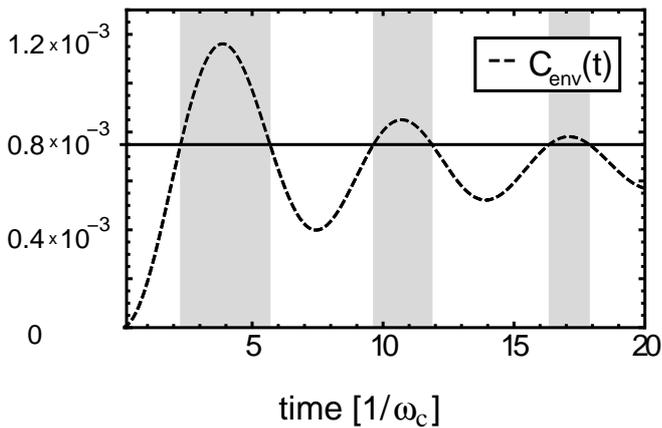}
  \caption{$C_\env (t)$ against time for a qubit with initial purity $r=0.997$. The highlighted domains correspond
to time intervals with non-zero system-environment entanglement. Clearly, an increase of
$C_\env (t)$ above a critical value indicates these quantum correlations.}
\label{abb:figure3}
\end{figure}

We can prove system-environment entanglement when the partial transpose $\rho_\tot^\PT$ of the total state yields a
negative expectation value $\bra{\Psi}\rho_\tot^\PT\ket{\Psi}$ in some
state $\ket{\Psi}$ of the composite system~\cite{peres-1996}. By means
of the representation~(\ref{equ:P-representation}) we have shown in~\cite{pernice+strunz-2011} that
with the time and temperature dependent function
\begin{equation}
\begin{split}
E(T,t)&=\;8\int_0^t\.ds\int_0^s\.d\tau\int_0^\infty\.\.d\omega\\
&\qquad\times\,J(\omega)\sinh\left[\hbar\omega/kT\right]\cos[\omega(s-\tau)]
\end{split}
\label{equ:E_t_continuum}
\end{equation}
entanglement is present whenever
\begin{equation}\label{equ:condition_ent}
E(T,t)>\ln\left[\frac{r-z^2}{x^2+y^2}\right].
\end{equation}
In contrast to the separable case studied in Fig. \ref{abb:figure2}, choosing a qubit
initial state with a purity closer to one, we can indeed prove the existence of
system-environment entanglement (see highlighted regions in Fig.~\ref{abb:figure3}). Remarkably, there
is a close connection between our ``entanglement witness''~(\ref{equ:condition_ent}) and the environmental contribution $C_\env(t)$ of the correlations.
In the high temperature limit $kT\gg \hbar\omega_c$ we find
\begin{equation}\label{equ:connection_E_Cenv}
\dot C_\env=\frac12\frac{\dot E}{b\,e^{-8E}+1},
\end{equation}
and for $E(T,t)\ll1$, $C_\env(t)=E(T,t)/2(b+1)$. Now we are able to reformulate the entanglement criterion (\ref{equ:condition_ent})
in terms of the environmental part of the system-environment correlations
\begin{equation}\label{equ:ent_crit_Cenv}
 C_\env(t)>\frac{\ln\left[\frac{r-z^2}{x^2+y^2}\right]}{2(b+1)}.
\end{equation}
In Fig.~\ref{abb:figure3} we choose a larger initial state purity ($r=0.997$), leading to time intervals with system-environment
entanglement (highlighted areas), according to criterion (\ref{equ:condition_ent}).
The appearance of quantum correlations is closely related to the dynamics of $C_\env(t)$ as explained earlier: in
Fig.~\ref{abb:figure3} values of $C_\env(t)$ larger then the threshold given by (\ref{equ:ent_crit_Cenv}) indicate entanglement.
Comparing these findings to the domains of negative $\gamma$ from Fig.~\ref{abb:figure2} (which are independent of the qubit initial state), we see no obvious 
relation between time intervals of \emph{quantum} correlations and non-Markovianity of the dynamics.

\section{Conclusions}\label{sec7}
We have investigated non-Markovian dynamics of a decohering qubit and its environment. Since the reduced dynamics can be modeled
by means of random unitary evolution, we have stressed that a genuine ``open quantum system'' point of view is not required in this case.
In particular, we argue that approaches based solely on the reduced 
description may be misleading with respect to interpretations. A study of ``information flow to the environment'', e.g., is questionable without the existence of environmental dynamical degrees of freedom.

Considering the full dynamics of system plus quantum environment, we have investigated the measure
$C=C_\sys+C_\env$ for system-environment correlations that emerge from an increase of the local entropies of the two subsystems. We have found (the time derivative of) this quantity to be closely related to the sign of the dephasing rate $\gamma(t)$, reflecting the non-Markovian character of the dynamics. Referring to earlier work,
we are able to show that the total state underlying the correlations described 
by $C$, is separable for a large class of mixed qubit initial states. 
Therefore, given the quantum environment, ``non-Markovianity'' is still unrelated to the build-up or decay 
of \emph{quantum} correlations (entanglement) between system and environment.

For qubit initial states with large purity, by contrast, we were able to find 
periods where the total state is entangled. But again, we see no
obvious relation between ``non-Markovianity'' and the build-up or decay of 
these quantum correlations. Interestingly though, we are able to relate 
the environmental part $C_\env$ of the correlations $C$ to entanglement.

We are confident that our approach will be helpful for further 
investigations with respect to system-environment correlations and 
``information flow'' in open system dynamics.

\section*{Acknowledgements}
A.P. and J.H. acknowledge support from the International Max Planck Research School 
(IMPRS) Dresden.

\end{document}